\documentclass[12pt]{article}		
\usepackage[super,sort&compress,comma]{natbib}

\usepackage{fancyhdr}		

\usepackage[section]{placeins}   %

\usepackage{graphicx}

\makeatletter \renewcommand\@biblabel[1]{$^{#1}$} \makeatother
\newcommand{\cen}[1]{\begin{center} #1 \end{center}}

\usepackage[mathlines]{lineno}

\usepackage{hyperref}
\hypersetup{ colorlinks,
    citecolor=blue,
    filecolor=blue,
    linkcolor=blue,
    urlcolor=blue
}

\usepackage{xcolor}

\definecolor{gray}{rgb}{0.6,0.6,0.6}
\definecolor{red}{rgb}{0.85,0,0}
\definecolor{green}{rgb}{0,0.85,0}
\definecolor{blue}{rgb}{0,0,0.85}
\definecolor{beige}{rgb}{0.92,0.87,0.78}
\usepackage[all]{hypcap}    

\newcommand{\be}{\begin{equation}} 
\newcommand{\ee}{\end{equation}}   

\begin{document}

\cen{\sf {\Large {\bfseries Improving the efficiency of small animal 3D printed compensator IMRT with beamlet intensity total variation regularization } \\  
\vspace*{10mm}
{\normalsize Xinmin Liu, PhD$^1$,
Alexander L. Van Slyke, PhD$^1$,
Erik Pearson, PhD$^2$,\\
Khayrullo Shoniyozov, PhD$^1$,
Gage Redler, PhD$^3$,
Rodney D. Wiersma, PhD$^1$}
} \\
\vspace*{6mm}
$^1$ Department of Radiation Oncology, University of Pennsylvania, Philadelphia, PA\\
$^2$ Department of Radiation and Cellular Oncology, University of Chicago, Chicago, IL\\
$^3$ Moffitt Cancer Center, Department of Radiation Oncology, Tampa, FL
\vspace{6mm}\\
\today\\
}

\pagenumbering{roman}
\setcounter{page}{1}
\pagestyle{plain}
Email: Xinmin.Liu1, Rodney.Wiersma@pennmedicine.upenn.edu\\

\begin{abstract}
\noindent {\bf Purpose:}
There is growing interest in the use of modern 3D printing technology to implement intensity-modulated radiation therapy (IMRT) on the preclinical scale which is analogous to clinical IMRT. However, current 3D-printed IMRT methods suffer from complex modulation patterns leading to long delivery times, excess filament usage, and inaccurate compensator fabrication. In this work, we have developed a total variation regularization (TVR) approach to address these issues.\\
{\bf Methods:}
TVR-IMRT, a technique designed to minimize the intensity difference between neighboring beamlets, was used to optimize the beamlet intensity map, which was then converted to corresponding compensator thicknesses in copper-doped PLA filament. IMRT and TVR-IMRT plans using five beams were generated to treat a mouse heart while sparing lung tissue. The individual field doses and composite dose were delivered to film and compared to the corresponding planned doses using gamma analysis.\\
{\bf Results:}
TVR-IMRT reduced the total variation of both the beamlet intensities and compensator thicknesses by around 50\% when compared to standard 3D printed compensator IMRT. The total mass of compensator material consumed and radiation beam-on time were reduced by 20-30\%, while DVHs remained comparable. Gamma analysis passing rate with 3\%/0.3mm criterion was 89.07\% for IMRT and 95.37\% for TVR-IMRT. \\
{\bf Conclusions:} TVR can be applied to small animal IMRT beamlet intensities in order to produce fluence maps and subsequent 3D-printed compensator patterns with less total variation, simplifying 3D printing and reducing the amount of filament required. The TVR-IMRT plan required less beam-on time while maintaining the dose conformity when compared to a traditional IMRT plan.\\

\end{abstract}

\newpage     

\tableofcontents

\newpage

\setlength{\baselineskip}{0.7cm}      

\pagenumbering{arabic}
\setcounter{page}{1}
\pagestyle{fancy}
\section{Introduction}

Intensity-modulated radiation therapy (IMRT) has become one of the predominant forms of radiation therapy (RT) currently utilized in clinical treatments. This prevalence is the result of ongoing research efforts that continue to improve the conformity and accuracy with which IMRT plans can be delivered. As the precision of equipment and accuracy of treatment planning algorithms improves, understanding the radiobiology involved becomes increasingly important \cite{rosenthal_mouse_2007,liu_lessons_2013,butterworth_small_2015,coleman_improving_2016}. As such, preclinical studies are essential for fully investigating novel RT treatment techniques and technologies before they can be implemented in the clinic \cite{butterworth_small_2015}. Accurate delivery of radiation to the site of disease in humans is a critical part of achieving a favorable outcome, so it would reasonably follow that the analogously accurate delivery of radiation to small animals desireable in order to facilitate translation of preclinical findings to clinical improvements. Given the small scales involved in mice, the tolerance of spatial deviation is on the submillimeter scale, rather than the multiple millimeter tolerance acceptable in humans \cite{desrosiers_use_2003}. For these reasons, the need for IMRT and image guided RT in preclinical studies at this submillimeter scale is important for enhancing the clinical relevance of associated results.

Currently, small animal IMRT methods include the use of millimeter scale secondary collimators with standard geometric shapes, and 3D-printed compensators to create a specific fluence map.
Secondary collimators has been implemented in two forms, the first of which utilizes a series of motorized jaw positions to create small rectangular segments delivered in a serial fashion and the second uses a Sparse Orthogonal Collimator (SOC), which consists of 4 sets of orthogonally positioned leaves \cite{stewart2013two,nguyen2016novel,reinhart_uimrt_2018,woods2019sparse1,woods2019sparse2}. Aperture-based methods are limited by a minimum size of the apertures, as well as potentially long treatment times requiring numerous exposures through different jaw/leaf positions. Another approch is through compensator-based methods which utilize 3D printed patterns through which the beam is delivered at specified angles. In order to achieve their attenuation, a shaped polylactic acid (PLA) compensator can either be packed with sodium iodide powder, or the compensator itself can be printed out of PLA impregnated with copper or tungsten \cite{yoon2020method,gage2020small}. Some advantages of the compensator method are the submillimeter resolution achievable by modern 3D printers, a reduction in the number of segments that need to be delivered in order to achieve a conformal dose, and the absence of additional mechanical parts which require maintenance and quality control \cite{stewart2013two,gage2020small}. However, 3D-printed compensators are specific to each plan and beam angle, thus a new set needs to be made for each plan, consuming additional material and printing time. The printing process, while capable of the aforementioned submillimeter accuracy, is more prone to error in abutting regions of greatly differing attenuation (and therefore thickness). Furthermore, adjacent regions with a large difference in intensity will lose lateral electronic equilibrium, as electrons scatter out of the copper in the highly attenuated area into the column of air above low-attenuation segments. 

Total variation regularization (TVR) is most often used in digital image processing for noise removal. Total variation refers to the integral of the absolute gradient of the signal. TVR is based on the principle that noisy signals have high total variation \cite{rudin1992nonlinear}.TVR has been adopted in the field of radiation oncology in the form of compressed sensing used during inverse treatment planning\cite{zhu2009search}. Compressed sensing facilitates planning by reducing the complexity of the optimized fluence map, reducing the number of multi-leaf collimator segments for a given treatment plan without sacrificing plan quality. A TVR term can be added to the cost function in order to minimize the complexity of fluence patterns for IMRT segments, or in this case, modulation between adjacent beamlets.

In this work we develop a TVR-based inverse planning method for IMRT in small animals based on a novel micrometer-scale dose deposition kernel designed for accurate treatment planning at the sub-millimeter scale. This approach is designed to reduce the complexity of the modulation pattern, overall beam-on time, and the amount of material used to create 3D printed compensators, while maintaining or improving the DVH that can been achieved with previous compensator-based IMRT methods.

\section{Methods and materials}

\subsection{TVR-based inverse planning}

In TVR based inverse planning for small animal, a gradient operator is defined as
\be\label{equ_xuv}
\nabla x(u,v) = \bigl|x_{u,v} - x_{u-1,v}\bigl| \;+  \; \bigl|x_{u,v} - x_{u,v-1}\bigl|,
\ee
where $x$ is beamlet intensity, and $u$ and $v$ are its row and column indices, respectively.
The total variation (TV) of intensity is defined as the
summation of the gradients in beam view of all fields as the following  $\ell_1$ norm term,
\be\label{intensity_TV}
\phi=\sum_{u,v}  \nabla x(u,v)=\|Gx\|_1.
\ee
The complexity of the fluence map can be evaluated by TV of intensity.
The TVR based IMRT includes this extra gradient term, and so it is given by
\be\label{form_compressed}
\begin{array}{rl}
\mbox{minimize} & h(y) + \beta \|Gx\|_1 \\
\mbox{subject to}
 & y=Dx,\;\; 0\preceq x \preceq u_x,
\end{array}
\ee
where  
$D$ is the dose matrix,
$y$ is the dose to the voxels,
and $u_x$ is the upper bounds of beamlet intensity,
$\beta$ is the weight of variation.
The function $h(y)$ is the cost function for conventional IMRT, and it can be a linear, quadratic or other nonlinear functions of the voxel doses of the targets and organs at risk (OAR).
Due to the  $\ell_1$ norm term, the TVR-IMRT cost function in (\ref{form_compressed}) is now convex but not differentiable.
To apply the efficient 
Limited-memory Broyden–Fletcher–Goldfarb–Shanno optimization algorithm (LBFGS), in which the derivative of the objective function is used, it is therefore necessary to approximate $|x|$ by a smooth function. The function $|x|$ can be approximated by $\sqrt{x^2+\nu^2}$ as in \cite{ramirez2014x2}, or by parabolic functions at the neigherhood of $x=0$. The differentials involved in optimization can be derived accordingly.

\subsection{Development and 3D Printing of Compensators}

\begin{figure}[ht]
\centerline{
		\includegraphics[height=73mm]{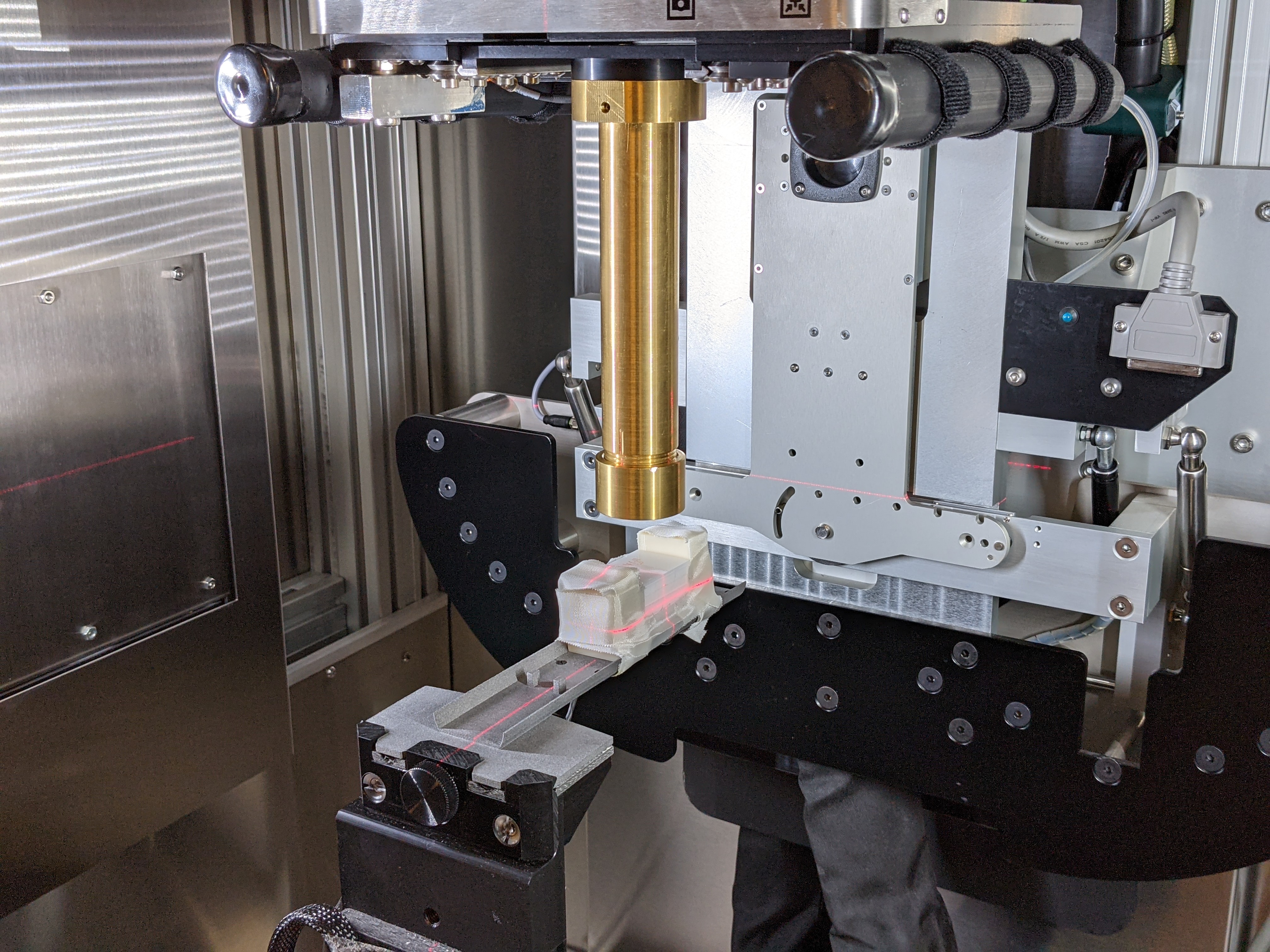}\hspace{10mm}
\begin{minipage}[b]{0.3\textwidth}
	\centerline{
		\includegraphics[width=65mm]{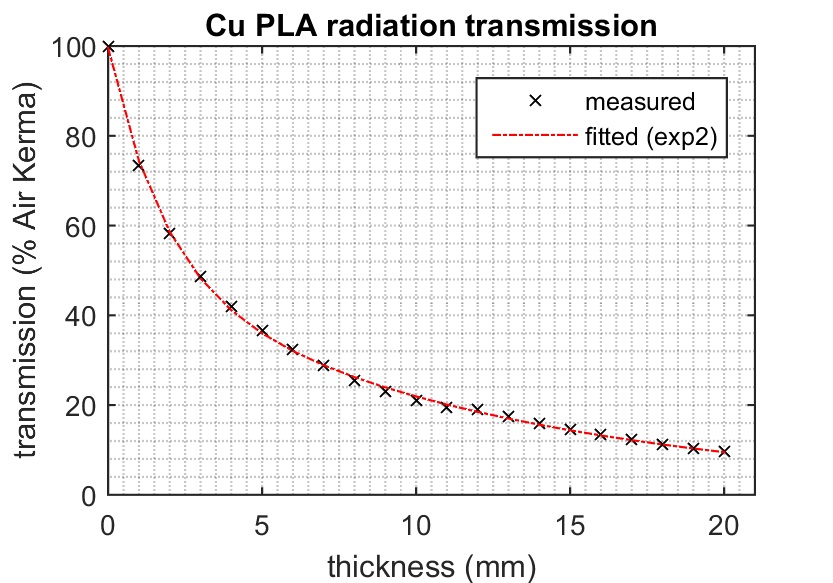}
		}\vspace{3mm}
	\centerline{
		\includegraphics[width=54mm]{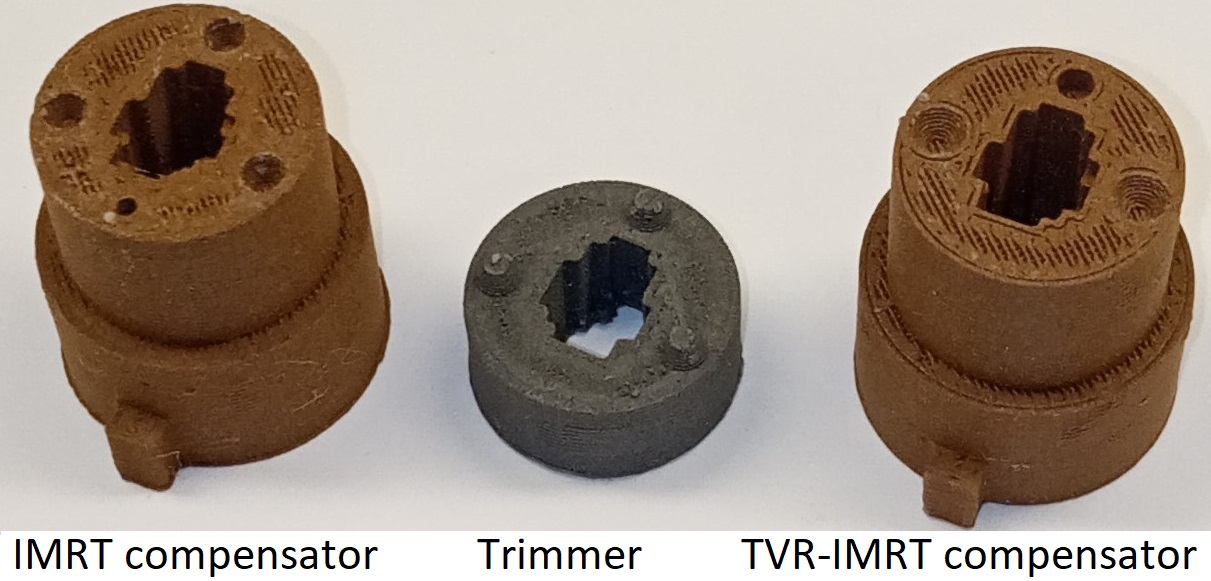} 
		}
\end{minipage}
}
	\caption{Left panel: Irradiation and measurement setup. Right upper panel: Radiation transmission of copper-PLA. Right lower panel: 3D-printed tungsten-PLA trimmer (black) and corresponding copper-PLA IMRT and TVR-IMRT compensators.}\
	\label{SARRP-transmission}
\end{figure}

A photon beam passing through a sample of
thickness $z$ is attenuated via numerous types of interactions with the surrounding medium. The photons are transmitted according to
Beer–Lambert’s law \cite{kucuk2012determining}.
The parameters of polychromatic transmission through both copper- and tungsten-PLA filaments were calculated by measuring the radiation transmitted through material thicknesses ranging from 0 to 20 mm with an ionization chamber. See Figure~\ref{SARRP-transmission} in the next section for the copper-PLA transmission. The radiation attenuation of tungsten-PLA is about four times larger than copper-PLA. The copper-PLA was used to shape the fluence map inside the beams-eye-view region containing the target, while the tungsten-PLA was used to print the trimmers to block the radiation outside the desired field (analogous to portions of beam collimated by MLCs and jaws, respectively, with cinical treatment machines). 
In the case of our copper filament compensators, a thin base layer of 0.4mm is first printed in the modulation area, and the maximum thickness of the compensators was assumed to be 20mm.
Therefore, the bound of beamlet intensity $x$ in (\ref{form_compressed}) should be replaced by $l_x\preceq x \preceq u_x,$, and the ratio of minimum intensity $l_x$ to maximum intensity $u_x$ for the optimization was set to 
\be\label{l_x_u_x}
l_x/u_x=T_{\tiny\mbox{Cu}}(20)/T_{\tiny\mbox{Cu}}(0.4),
\ee
where $T_{\tiny\mbox{Cu}}(z)$ is the transmission through Copper-PLA filament of thickness $z$.
The optimization was first run without constraint on fluence intensity, then $u_x$ was set to an appropriate value accordingly.
Once obtaining the optimized bixel intensity $x$, the compensator thickness can be computed accordingly. Denote $x=\mbox{col}(x_1,\cdots,x_m)$, where $m$ is number of gantry angles. For the $i$-th each gantry angle,
the expose time $t_i=\mbox{max}(x_i/T_{\tiny\mbox{Cu}}(0.4))$. The dose in a single time interval is $x_i/t_i$, and thus the thickness $z_i$ is given by
\[
z_i=T_{\tiny\mbox{Cu}}^{-1}(x_i/t_i),
\]
The total variation (TV) of thickness is defined in the similar way as in (\ref{equ_xuv}) and (\ref{intensity_TV}),
\be\label{thickness_TV}
\varphi=\sum_{u,v}  \nabla z(u,v)=\|Gz\|_1,
\ee
which can be used to evaluate the complexity of 3D printing surface inside beamview.

The compensators used in this study were printed using a Raise3D Pro2 Plus (Raise3D Technologies Inc, Irvine, CA). The current version of these 3D-printed compensators have a tab and outer ring for mounting and alignment. The associated tungsten trimmer is positioned downstream, in contact with the compensator, in order to remove any scatter from outside the desired region. See Figure~\ref{SARRP-transmission}.

\subsection{IMRT plan for a mouse heart with lung OARs}

A comprehensive practical implementation of IMRT using 3D-printed beam compensators on the Xstrahl Small Animal Radiation Research Platform (Xstrahl Inc, Suwanee, GA) (SARRP) was used to illustrate the TVR method. 
The Source-Axis-Distance (SAD) of the SARRP is 350mm, and, for the customized collimator, the source-to-compensator distance is 290mm. The bixel width was set to 1.0mm at SAD for all treatment planning, thus the bixel width of the 3D-printed compensators was 290/350=0.8286mm.

Plans were generated from a CT taken for use in a study on heart toxicity, using the built-in imaging system on the SARRP \cite{dreyfuss_novel_2021}. Hounsfield unit (HU) values were discretized to 5 categories: air, lung, fat, muscle, and bone, which were assigned an electron density by Muriplan (xStrahl) which are based on ICRU-44\cite{white_report_1989}. The CT dimensions used for optimization were $126\times 126\times 134$ voxels with a voxel width of 0.254mm. The lengths of target in $xyz$ direction are 8.9, 5.6, and 8.6mm. Relative dosimetry measurements and subsequent dose calculation using the 3D printed compensators was verified as in the previous work \cite{redler2020small}.

The heart was designated as the planning target volume (PTV) and the lungs were the only organ at risk (OAR) contoured and given dose constraints. 
The plan generated by each optimization method was evaluated by comparing the $D95_{\tiny\mbox{PTV}}$[\%], $V95_{\tiny\mbox{PTV}}$[\%], $D95_{\tiny\mbox{lung}}$[\%], $V95_{\tiny\mbox{lung}}$[\%], and $V110_{\tiny\mbox{lung}}$[\%], as well as the total variation (TV) of both the fluence and compensator thicknesses, and total beam-on time.

In order to evaluate the accuracy of the delivered plan, each of the five beams, as well as the composite dose was delivered to EBT2 film placed in a $32\times 32$ mm high-density polyethylene (HDPE) phantom with the center of the film placed at isocenter. The accuracy of the dose calculation algorithm and resultant compensator was assessed both on a per-field and composite dose planar gamma analysis by using previously published software \cite{low1998technique} \cite{mark2015gamma}.

\section{Results}

\subsection{Experimental setup and copper filament transmission}

For this study, a custom brass collimator and compensator-holder was machined to fit the Xstrahl SARRP. This collimator held a series of notches and alignment pins to ensure that the compensator for each beam was held at the same source to compensator distance, without allowing rotation around the beam axis. The $32\times 32$ mm HDPE phantom can be seen with the film inside placed at isocenter (Figure 1, left panel).

Transmission through a range of Cu-PLA thicknesses was measured in order to translate desired bixel intensity into a physical compensator (Figure 1, right upper panel). Cu-PLA compensators for each beam angle were designed with and without TVR. A tungsten-PLA trimmer was designed to reduce out of field scatter and aligned to the compensators with three tongue-and-groove points (Figure 1, right lower panel).

\subsection{Plan delivery statistics and compensator dimensions}
\begin{figure}[ht]
	\centerline{
		\includegraphics[width=50mm]{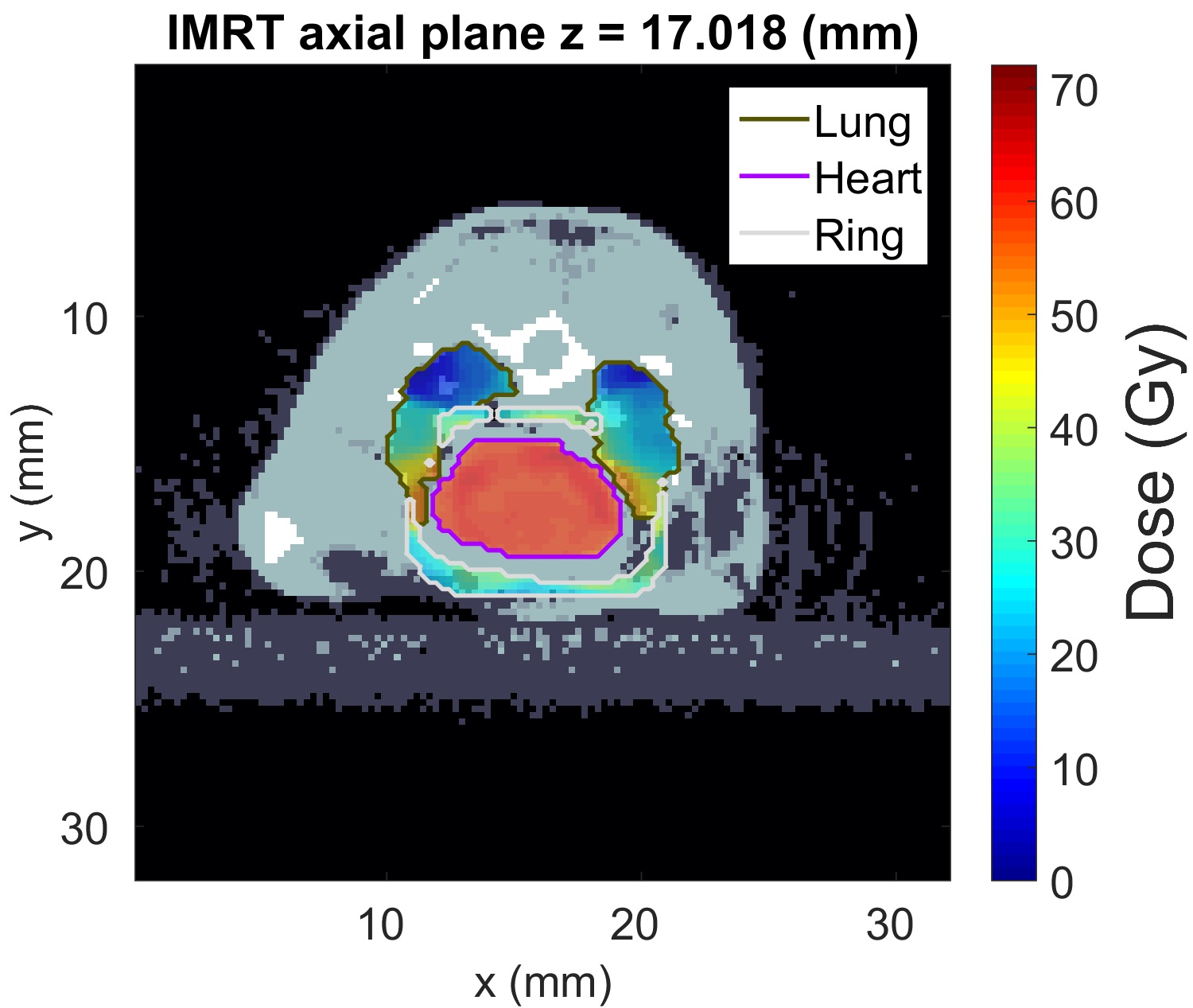}\hspace{-9.7mm}
		\includegraphics[width=50mm]{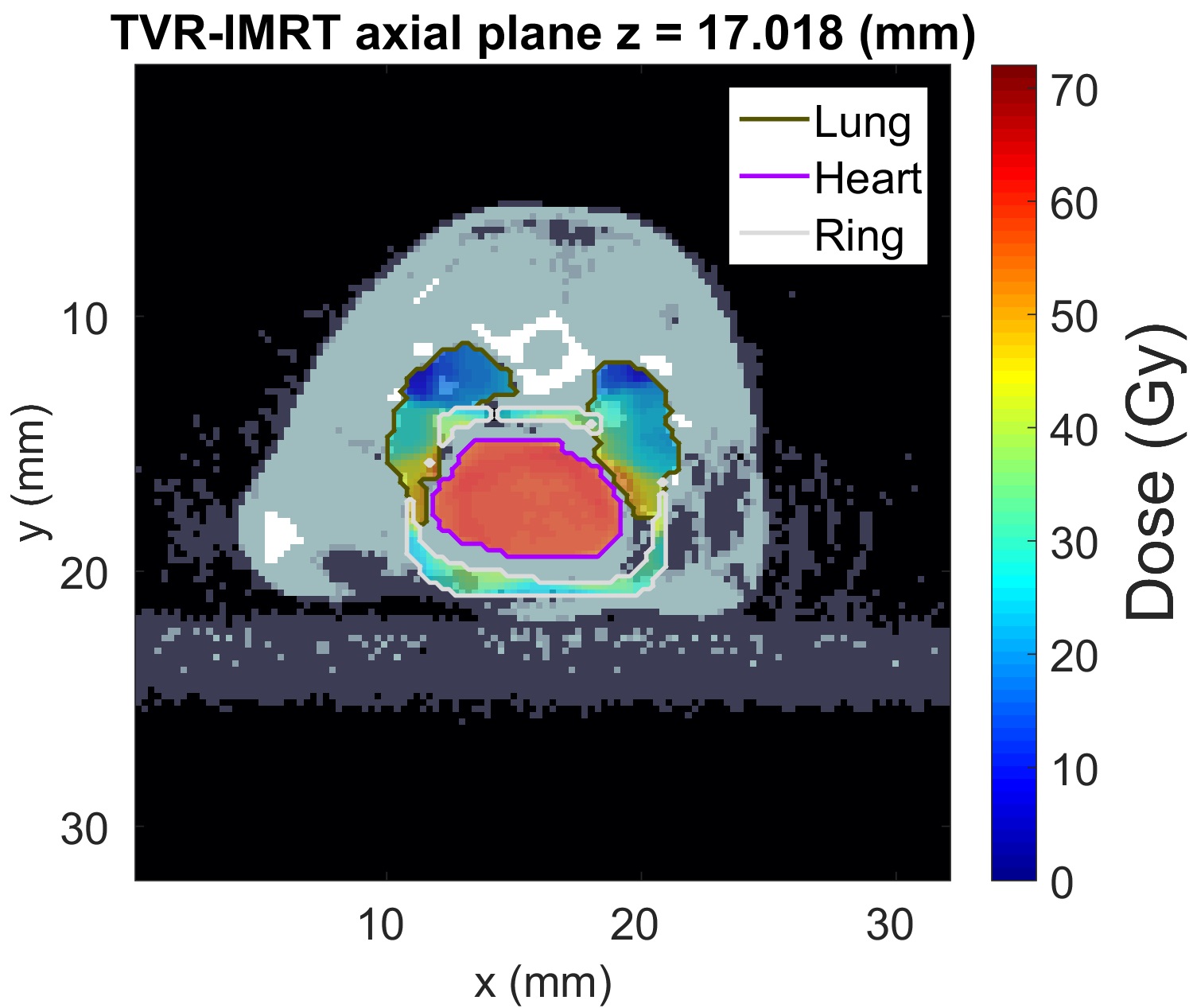}\hspace{6mm}
		\includegraphics[width=60mm]{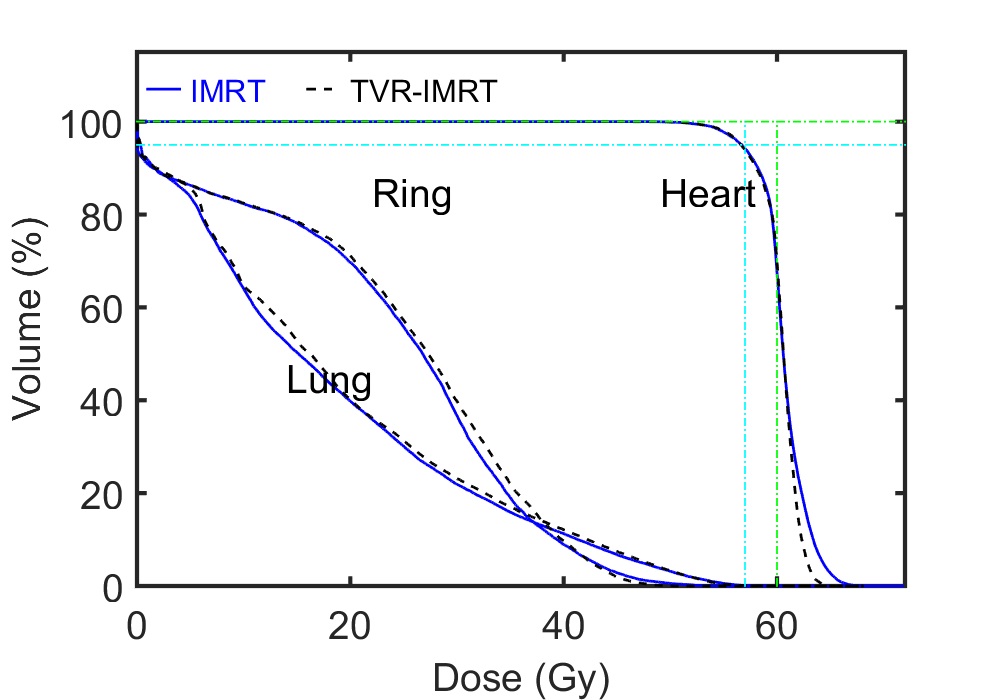}
	}
	\caption{Dose distribution and DVH comparison for IMRT (solid) and TVR-IMRT (dashed)
	}
	\label{fig_Dose}
\end{figure}

The dose heat maps shown on an axial slice of the mouse CT (Figure~\ref{fig_Dose}) containing the heart and lungs and the corresponding DVH were similar between both planning methods,
with the TVR-IMRT having slightly more homogenity (lower hotspot). This is also shown by the data in Table~\ref{table_D95V95}.

\begin{table}[htbp]
\caption{DVH statistics comparison of optimization results}
\centerline{\small
\begin{tabular}{l|rrr|rrr}
  \hline\hline
&  PTV $D_{95}$ [\%] &  $V_{95}$ [\%]   &$V_{110}$ [\%]   & Lung $D_{95}$ [\%] &  $V_{50}$ [\%] & $V_{95}$ [\%] \cr
\hline
      IMRT &     94.5\%&     94.2\%&  0.9\%&   0.0\%&     21.9\%&      0.1\% \cr
TVR-IMRT&     94.4\%&     93.9\%&  0.0\%&   0.1\%&     23.1\%&      0.1\% \cr
\hline\hline
\end{tabular}
}
\label{table_D95V95}
\end{table}

All dose objectives were met by both plans and were within 0.5\% for all but the PTV $V_{110}$ and the Lung $V_{50}$ (Table 1). The TVR-IMRT plan did not deliver 110\% of the dose to any portion of the PTV, while the IMRT plan delivered it to 0.9\%. The Lung $V_{50}$ was 1.2\% lower for the IMRT plan than the TVR-IMRT plan (Table~\ref{table_D95V95}).

\begin{figure}[ht]
	\centerline{\includegraphics[width=170mm]{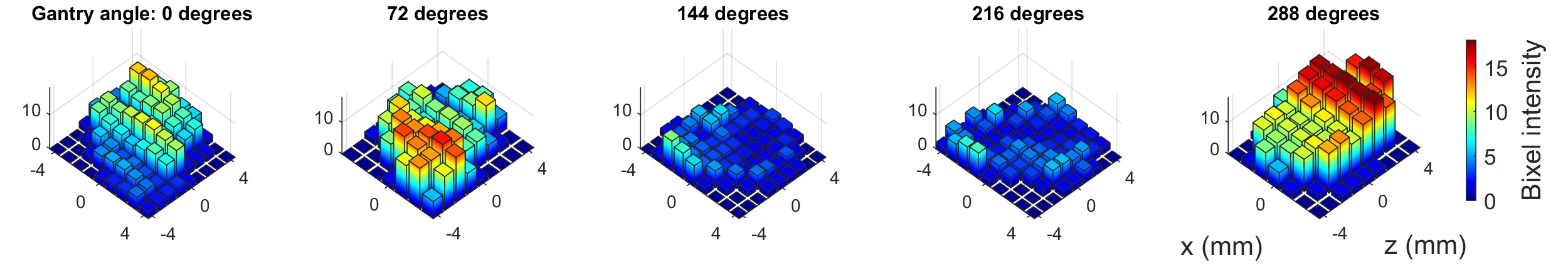}}
	\vspace{2mm}
	\centerline{\includegraphics[width=170mm]{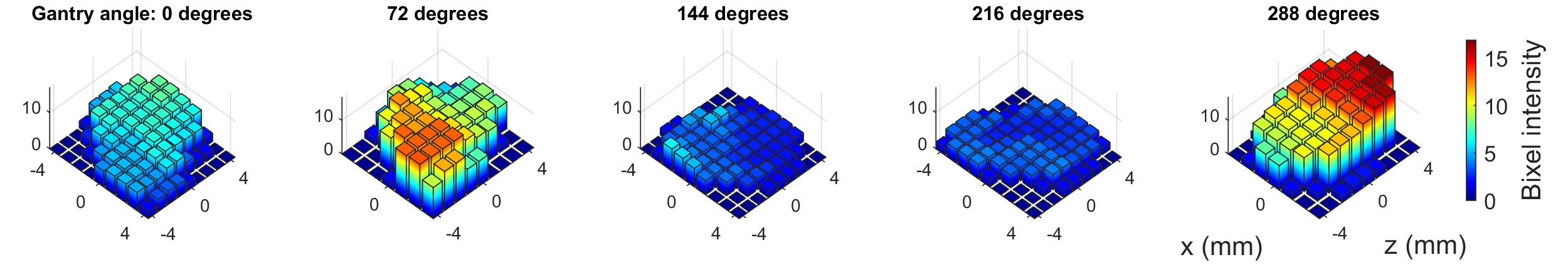}}
	\caption{Bixel intensity comparison of the compensators designed for each gantry angle. Top: IMRT. Bottom: TVR-IMRT}
	\label{fig_beamlet_intensity}
\end{figure}

\begin{figure}[ht]
	\centerline{\includegraphics[width=170mm]{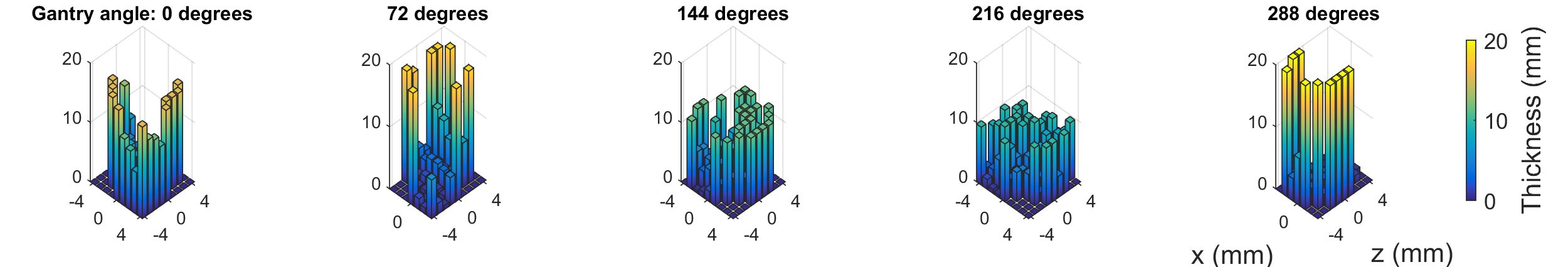}}
	\vspace{2mm}
	\centerline{\includegraphics[width=170mm]{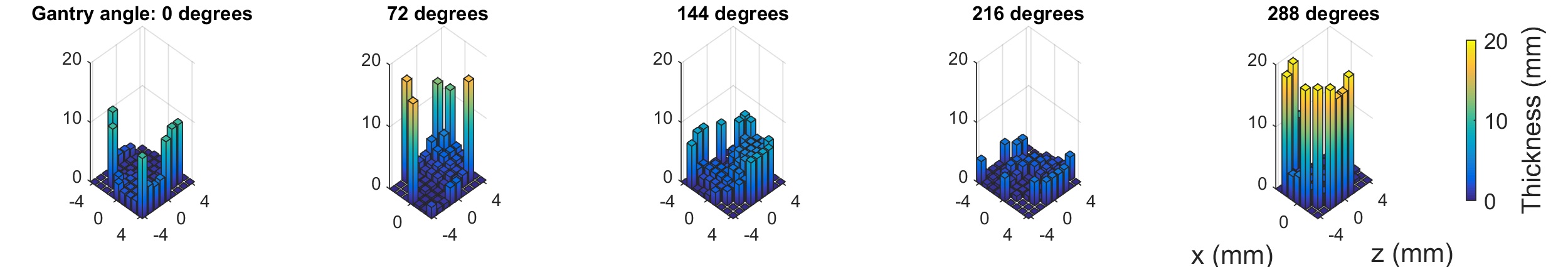}}
	\caption{Compensator thicknesses for each gantry angle.  Top: IMRT. Bottom: TVR-IMRT}
	\label{fig_compensators_thick}
\end{figure}

\begin{table}[htbp]
	\caption{IMRT vs TVR-IMRT}
	{\small
		\begin{tabular}{l|r|r|r|r}
			\hline\hline
			&  Intensity TV & Thickness TV (mm) & Total thickness (mm) & Exposure time (s) \cr
			\hline
			IMRT &      997 (   0\%)&     1620 (   0\%)&     1771 (   0\%)&   280 (   0\%) \cr
			TVR-IMRT&      517 ( -48\%)&     728 ( -54\%)&     902 ( -28\%)&   218 ( -22\%) \cr
			\hline\hline
		\end{tabular}
	}
	\label{table_compare_TV}
\end{table}

The TV of the delivered dose and physical thickness for the TVR-IMRT compensator was reduced by 48\% and 54\% respectively, when compared to the IMRT compensator as shown in Table~\ref{table_compare_TV}. TVR reduced the sum of the thicknesses of each discrete region in the beam path (corresponding to the filament used during printing) by 28\%. The time to deliver the plan was similarly decreased by 22\% for TVR-IMRT.

\subsection{Gamma analysis}

\begin{figure}[ht!]
	\centerline{
	\includegraphics[width=160mm]{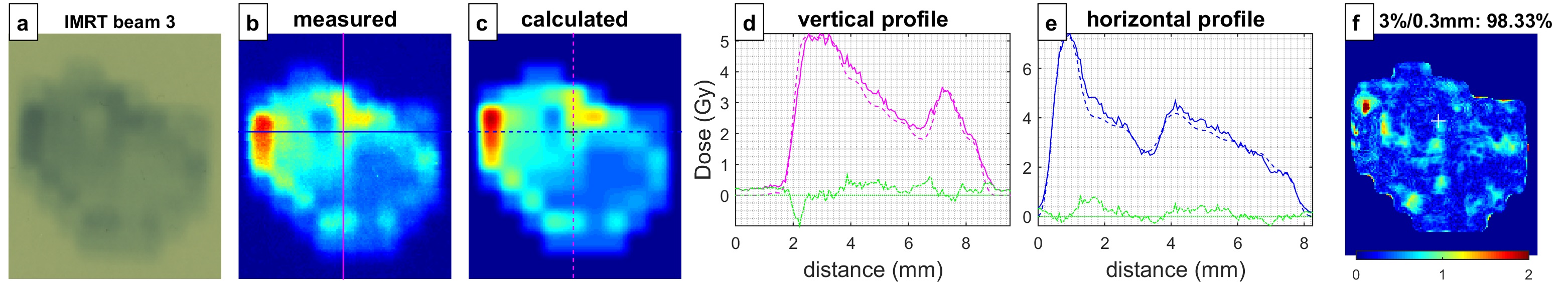}}
	\vspace{2mm}
	\centerline{
	\includegraphics[width=160mm]{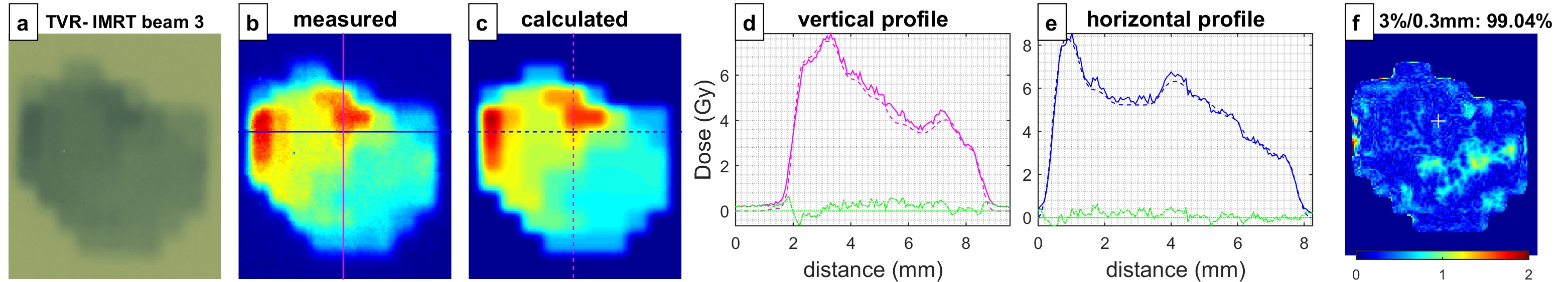}}
	\caption{Gamma analysis of single field dose delivery. a: films. b and c: measured and calculated dose heat maps.
			d and e: profiles for measured dose (dotted line) and calculated dose (solid line). 
			Green line denotes difference between measured and calculated.
			f: Gamma analysis of 3\%/0.3mm heat maps (Top: IMRT. Bottom: TVR-IMRT) 
			}
	\label{fig_gamma1}
\end{figure}

\begin{figure}[ht!]
	\centerline{
	\includegraphics[width=160mm]{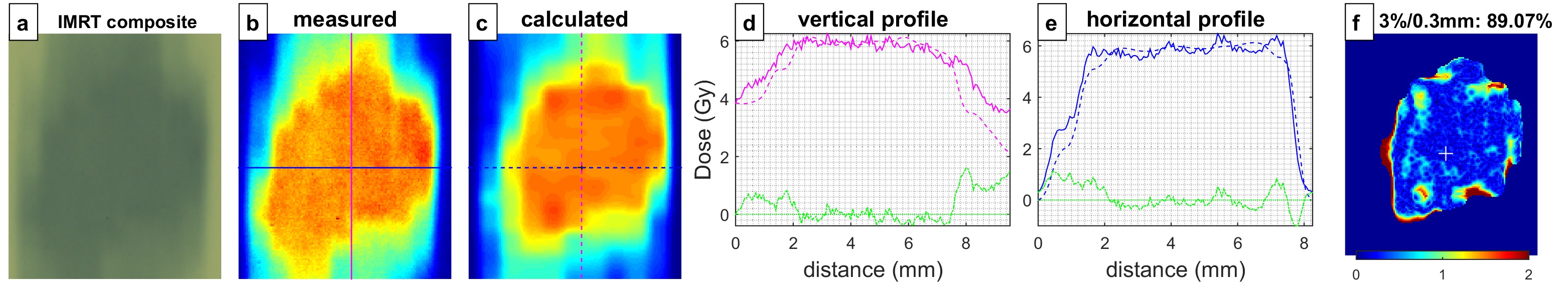}}
	\vspace{2mm}
	\centerline{
	\includegraphics[width=160mm]{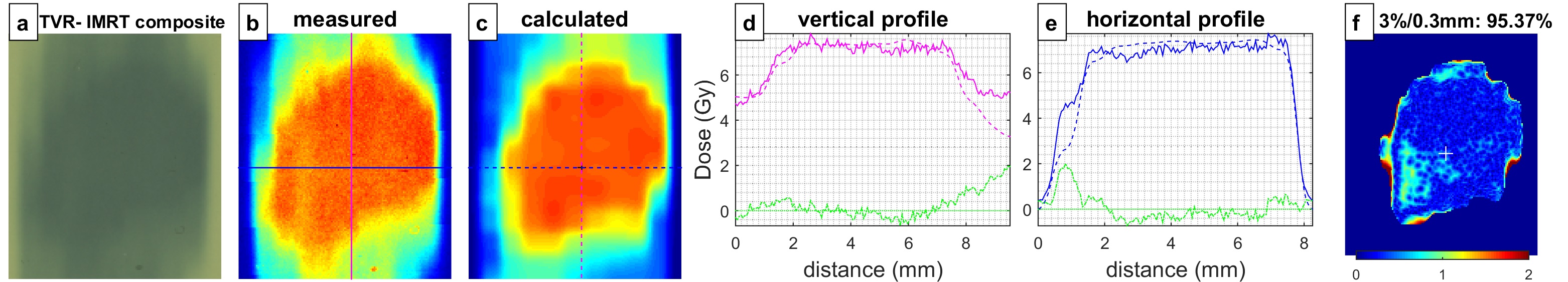}}
	\caption{Gamma analysis of composite dose delivery. Each column is the same as that in Figure~\ref{fig_gamma1}. (Top: IMRT. Bottom: TVR-IMRT) 
			}
	\label{fig_gamma2}
\end{figure}

The planar dose profile (perpendicular to beam-axis) of each individual compensator for each of 5 gantry angles (0$^\circ$, 72$^\circ$, 144$^\circ$, 216$^\circ$, 288$^\circ$), 
were all delivered from gantry angle 0$^\circ$ to a HDPE phantom for ease of comparison and to facilitate gamma analysis for both IMRT and TVR-IMRT (Figure~\ref{fig_gamma1}). 
Composite dose delivery with the film horizontally placed for each plan was also performed (Figure~\ref{fig_gamma2}).

The films were calibrated by net-OD with actual dose delivered to the films, and fitted by a polynomial with degree of two. The measured dose distribution was compared with calculated dose distribution by dose profiles and gamma analysis.

For single field dose delivery (Figure~\ref{fig_gamma1}), the measured and calculated dose distributions match each other very well, which indicated that all dose calculation algorithm, measurement of radiation transmission, 3D printing, experimental dose delivery and film calibration worked appropriately. Compared to IMRT, the TVR-IMRT plan appears more blurred, which is evidence of the reduction in fluence TV. Gamma analysis indicated that TVR-IMRT has a slightly better fidelity than IMRT.


The gamma analysis for each plan composite dose delivery with 3\%/0.3mm criterion achieved a passing rate of 89.07\% and 95.37\% for IMRT and TVR-IMRT respectively, but the IMRT plan showed more distinct regions with significant deviation around the edges of the field (Figure~\ref{fig_gamma2}).

The gamma analysis of the composite dose meets the 90\% clinical passing threshold at 3\%/0.5mm for the IMRT plan, while TVR-IMRT passes at the more strict 3\%/0.2mm criteria (Table 3).
\begin{table}[ht]
	\caption{Gamma analysis results for IMRT and TVR-IMRT}
	{\small
		\begin{tabular}{r|r|r|r|r|r|r}
			\hline\hline
			  & beam1 & beam2 & beam3 & beam4 & beam5 & composite\cr
			\hline
			3\%/0.5mm: $\qquad$ IMRT&     99.60&    99.15&    99.69&    97.88&    96.59&    94.35\cr
			TVR-IMRT&        99.53&    98.81&    99.79&    99.52&    97.84&    97.51\cr
			\hline
			3\%/0.3mm: $\qquad$ IMRT& 99.23&    97.36&    98.33&    93.62&    94.05&    89.07    \cr
			TVR-IMRT&    99.09&    97.76&    99.04&    96.63&    95.28&    95.37         \cr
			\hline
			3\%/0.2mm: $\qquad$ IMRT&     97.51&    94.03&    93.80&    86.23&    87.76&    84.11    \cr
			TVR-IMRT&        98.09&    93.47&    96.79&    88.78&    88.96&    91.42    \cr
			\hline\hline
		\end{tabular}
	}
	\label{table_compare_gamma}
\end{table}

\section{Discussion and Conclusions}

 Preclinical research is essential to the continued development of RT technologies, fractionation schema, and knowledge of the underlying radiobiology which have all contributed to the improvement of clinical outcomes. The growing interest in small animal IMRT stems, in part, from the high degree of conformality and precision that is demanded in the clinic. In order to best predict the biological effects and outcomes of a procedure prior to use in the clinic, preclinical studies must be performed which can adhere to those same standards. 

Dose painting and the associated sparing of neighboring healthy tissues has been proven to improve patient outcomes, so certain preclinical studies could benefit from these clinical-quality plans even though it would result in a smaller sample size. 

We have shown that this method of planning is capable of producing the high fidelity dose distributions which make 3D-printed compensators an attractive method of small animal IMRT, with the plan generated by both IMRT and TVR-IMRT meeting all dose constraints and remaining within 1.2\% of one another for all of the plan evaluation metrics that we used. This was achieved while limiting the total variation, resulting in compensator modulation patterns that have shallow physical gradients relative to previous design methods\cite{gage2020small,yoon2020method}. 

The reduction in steep gradients and variation produces a compensator design which enables higher fidelity printing, as shown by both the individual field dose heat maps and improved gamma passing rate for the TVR-IMRT plan (Figures~\ref{fig_gamma1}-\ref{fig_gamma2} and Table \ref{table_compare_gamma}). Furthermore, complex modulation patterns present in each beam angle may also result in a plan which is less robust with respect to small (sub degree, sub millimeter) variations in compensator positioning between dose deliveries, resulting in several areas of dose deviation in the composite dose films for the IMRT plan which were not present for the TVR-IMRT plan (Figure~\ref{fig_gamma2}). Both plans exhibited some deviation around the periphery, but it was more pronounced in the IMRT plan. The composite dose delivery of the TVR-IMRT plan passing the gamma analysis with 3\%/0.2mm criteria, compared with the IMRT plan only passing the 3\%/0.5mm criteria is further evidence that TVR-IMRT produces a compensator which is both more likely to print accurately and more robust to small positioning errors.

The TVR-IMRT method of 3D-printed compensator design was shown to provide an advantage over previous methods by reducing the sum of thicknesses used in the fluence map modulation region, resulting in lower cost and shorter print times, as well as reducing the total beam- on time for each exposure, while preserving or improving the dose distribution in the target as confirmed by DVH and gamma analysis.

In conclustion, here we have presented a technique utilizing total variation regularization (TVR) which improves upon the current design method for 3D-printed compensators for small animal IMRT. This is a subject which is garnering increasing interest due to the need to produce clinically analogous results in preclinical studies. TVR-IMRT limits the total variation across the modulated intensity pattern to be produced by the variable thickness compensator, resulting in more shallow gradients that facilitate high fidelity printing and can produce plans with improved gamma analysis passing rates. Furthermore, there is a significant reduction in total beam-on time to deliver a plan, total compensator thickness, 3D printer filament consumed, and print time when compared to previous methods.

\section*{References}
\addcontentsline{toc}{section}{\numberline{}References}
\vspace*{-15mm}

\bibliographystyle{ieeetr}    

\end{document}